\documentclass{article}
\usepackage{spconf,amsmath,graphicx}
\usepackage{amssymb,amsthm,mathrsfs,amsfonts}
\ninept

\title{A Learning Approach To Wireless Information and Power Transfer Signal and System Design}
\name{Morteza Varasteh, Enrico Piovano and Bruno Clerckx\thanks{This work has been partially supported by the EPSRC of the UK, under the grant EP/P003885/1.}}

\address{Department of Electrical and Electronic Engineering, Imperial College London, London, U.K.\\
\{m.varasteh12; e.piovano15; b.clerckx\}@imperial.ac.uk.}
\begin{document}
%
\maketitle
\begin{abstract}
The end-to-end learning of Simultaneous Wireless Information and Power Transfer (SWIPT) over a noisy channel is studied. Adopting a nonlinear model for the energy harvester (EH) at the receiver, a joint optimization of the transmitter and the receiver is implemented using Deep Neural Network (DNN)-based autoencoders. Modulation constellations for different levels of ``power'' and ``information rate'' demand at the receiver are obtained. The numerically optimized signal constellations are inline with the previous theoretical results. In particular, it is observed that as the receiver energy demand increases, all but one of the modulation symbols are concentrated around the origin and the other symbol is shot away from the origin along either the real or imaginary subchannel.
\end{abstract}
\section{Introduction}\label{Sec_Intro}
Radio Frequency (RF) signals are capable of bearing information as well as power. The transferred power can be utilized for energizing low power devices, such as wireless sensors and Internet-of-Things (IoT) devices. This along with the growth of low energy devices, has created a significant attention towards the study of Simultaneous Wireless Information and Power Transfer (SWIPT) systems \cite{Clerckx_Zhang_Schober_Wing_Kim_Vincent}. The fundamental tradeoff between the information rate and the delivered power was first studied in \cite{Varshney_2008} by Varshney, where a characterization of the capacity-power function for a point-to-point discrete memoryless channel is obtained.

In order to design efficient SWIPT architectures, it is crucial to model the energy harvester (EH) with a high level of accuracy. The EH consists of a rectenna, which is composed of an antenna followed by a rectifier. The rectifier is used to convert the RF power into DC current in order to charge devices. Although most of the results in the literature adopt a linear characteristic function for the rectifier, in practice, due to the presence of a diode in the rectifier, the output of the EH is a nonlinear function of its input \cite{Clerckx_Bayguzina_2016, Boaventura_Collado_Carvalho}.

Due to the nonlinearity of the diode characteristic function, the RF-to-DC conversion efficiency of the EH is highly dependent on the power as well as the shape of the waveform \cite{Clerckx_Bayguzina_2016,Boaventura_Collado_Carvalho,Clerckx_Bayguzina_2017}. Observations based on experimental results reveal that signals with high Peak-to-Average Power Ratio (PAPR) result in high delivered DC power compared to other signals \cite{Boaventura_Collado_Carvalho}. Motivated by this observation, in \cite{Clerckx_Bayguzina_2016}, an analytical model for the rectenna is introduced and a joint optimization over the phase and amplitude of a deterministic multisine signal is studied. It is concluded that unlike the linear EH model that favours a single-carrier transmission, a nonlinear model favours a multicarrier transmission.

In Wireless Power Transfer (WPT) systems, the goal is to design waveforms that maximize the DC power at the output of the EH, whereas, in SWIPT systems, the goal is to maximize the DC power as well as the information rate, which is commonly referred as maximizing the rate-power (RP) region. Unlike most of the SWIPT systems with the linear model assumption for EH, for SWIPT systems with nonlinear EH, there exists a tradeoff between the rate and delivered power \cite{Clerckx_Zhang_Schober_Wing_Kim_Vincent}. Due to the presence of nonlinear components in EH, obtaining the exact optimal tradeoff analytically has so far been unsuccessful. However, after making some simplifying assumptions, some interesting results have been derived in \cite{Clerckx_2016,Varasteh_Rassouli_Clerckx_ITW_2017,Morsi_Jamali,Varasteh_Rassouli_Clerckx_arxiv}. In particular, in multicarrier transmission, it is shown in \cite{Clerckx_2016} that nonzero mean Gaussian input distributions lead to an enlarged RP region compared to Circularly Symmetric Complex Gaussian (CSCG) input distributions. In single carrier transmissions over Additive White Gaussian Noise (AWGN) channel, in  \cite{Varasteh_Rassouli_Clerckx_arxiv,Morsi_Jamali}, it is shown that (under nonlinearity assumption for the EH) for circular symmetric inputs, the capacity achieving input distribution is discrete in amplitude with a finite number of mass-points and with a uniformly distributed independent phase. This is in contrast to the linear model assumption of the EH, where there is no tradeoff between the information and power (i.e., from system design perspective the two goals are aligned), and the optimal inputs are Gaussian distributed \cite{Clerckx_Zhang_Schober_Wing_Kim_Vincent}.

While designing SWIPT signals and systems (under nonlinear assumptions for the EH) using analytical tools seems extremely cumbersome, Deep Learning (DL)-based methods reveal a promising alternative to tackle the aforementioned problems. In fact, DL-based methods, and particularly, autoencoders have recently shown remarkable results in communications, achieving or even surpassing the performance of state-of-the-art algorithms \cite{OShea_Hoydis_2017,OShea_Karra_Clancy_2016}. The advantage of DL-based methods versus analytical algorithms lies in their ability to extract complex features from the training data, and the fact that their model parameters can be trained efficiently on large datasets via backpropagation. The DL-based methods learn the statistical characteristics from a large training dataset, and optimize the algorithm accordingly, without obtaining explicit analytical results. At the same time, the potential of DL has also been capitalized by researchers to design novel and efficient coding and modulation techniques in communications. In particular, the similarities between the autoencoder architecture and the digital communication systems have motivated significant research efforts in the direction of modelling end-to-end communication systems using the autoencoder architecture \cite{OShea_Hoydis_2017,OShea_Karra_Clancy_2016}. Some examples of such designs include decoder design for existing channel codes \cite{Nachmani_etall}, blind channel equalization \cite{Caciularu_Burshtein}, learning physical layer signal representation for SISO \cite{OShea_Hoydis_2017_2} and MIMO systems \cite{Timothy_OShea_Erpek}, OFDM systems \cite{Felix_Cammerer_Dorner,Ye_Li_Juang}.

\begin{figure}
\begin{centering}
\includegraphics[scale=0.43]{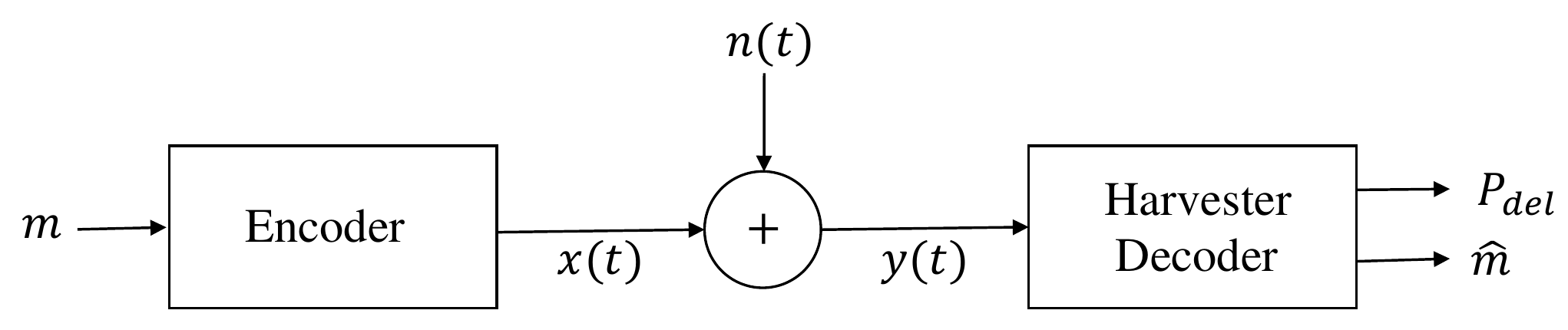}
\caption{Point-to-point SWIPT system model with an additive noise channel. The receiver is assumed to be capable of jointly capturing the power and decoding the information of the received signal.}\label{Fig_2}
\par\end{centering}
\vspace{0mm}
\end{figure}

In this work, we leverage DL-based methods in SWIPT. We consider signal modulation design for a point-to-point SWIPT over a noisy channel. In particular, we consider the SWIPT system as an autoencoder structure, where the transmitter and receiver are considered as multi-layer Deep Neural Networks (DNN). The results are obtained by optimizing the transmitter and receiver jointly over a large training data. The numerical optimization reveals the following: \textit{First}, as the demand for power at the receiver increases, one of the channel input symbols (namely, the power symbol) is getting away from zero with the other symbols (namely, the information symbols) distributed symmetrically around the origin. \textit{Second}, the power symbol is always along either the real or imaginary axis. This observation is inline with the result in \cite{Varasteh_Rassouli_Clerckx_ITW_2017}, where it is shown that as the power demand at the receiver increases, the transmitter allocates more power to either real or imaginary axis. It is also inline with the flash signaling interpretation of \cite{Varasteh_Rassouli_Clerckx_arxiv}. \textit{Third}, for higher power delivery demands, the number of channel information symbols decreases, i.e., the transmitter sacrifices some of the information symbols by mapping them to the same channel input (usually zero symbol). \textit{Fourth}, for power delivery purposes, the DC power increases with the number of channel input symbols, and all the symbols but one are with zero amplitude\footnote{We note that, in this paper, we focus on small-signal range analysis. Therefore, we have assumed to operate in the non-breakdown regime of the diode for reasons highlighted in \cite{Clerckx_2016}.}.

The rest of the paper is organized as follows. In Section \ref{Sec:sys_Model}, we introduce the system model, and provide some background on nonlinear EH. In Section \ref{Sec:Implemen}, we formulate the problem and introduce the DNN architecture. Section \ref{Sec:Numerical} is dedicated to the evaluation of the performance of the DNN architecture. Finally, Section \ref{Sec:Concl} concludes the work.

\section{System Model}\label{Sec:sys_Model}

The design of communication systems, in general, relies on the optimization of individual components of the transmitter and the receiver. However, in many scenarios, it is unclear whether this approach is the optimal possible design. Motivated by this, we aim at utilizing machine learning (ML) to enable optimization of SWIPT systems for end-to-end performance, without the need for dividing the transmitter and receiver into different sections.

We study a point-to-point SWIPT problem over an additive noise channel\footnote{In this paper, we consider Additive White Gaussian Noise (AWGN) for the channel noise, however, the approach can be extended to any noise model.}. The system model is shown in Figure \ref{Fig_2}, where the receiver is capable of harvesting the power (denoted by $P_{\text{del}}$) of the received signal as well as decoding the information, jointly\footnote{The tools presented in this paper can be easily extended to the scenario where there is a power splitter at the receiver as in \cite{Clerckx_2016}.}. The baseband information bearing pulse modulated signal is represented as $x(t) = \sum_{k=-\infty}^{\infty}x[k]g(t-kT)$, where $g(t)$ is the pulse waveform and $x[k]$ is the k$^{\text{th}}$ complex information-power symbol at time $k$. The received signal in the baseband is $y(t) = x(t)+n(t)$, where $n(t)$ is the baseband complex-valued noise. The EH is fed with the received RF signal, i.e., $y_{\text{RF}}(t)=\sqrt{2}\text{Re}\{y(t)e^{j2\pi f_c t}\}$, where $f_c$ is the carrier frequency.

\begin{figure}
\begin{centering}
\includegraphics[scale=0.45]{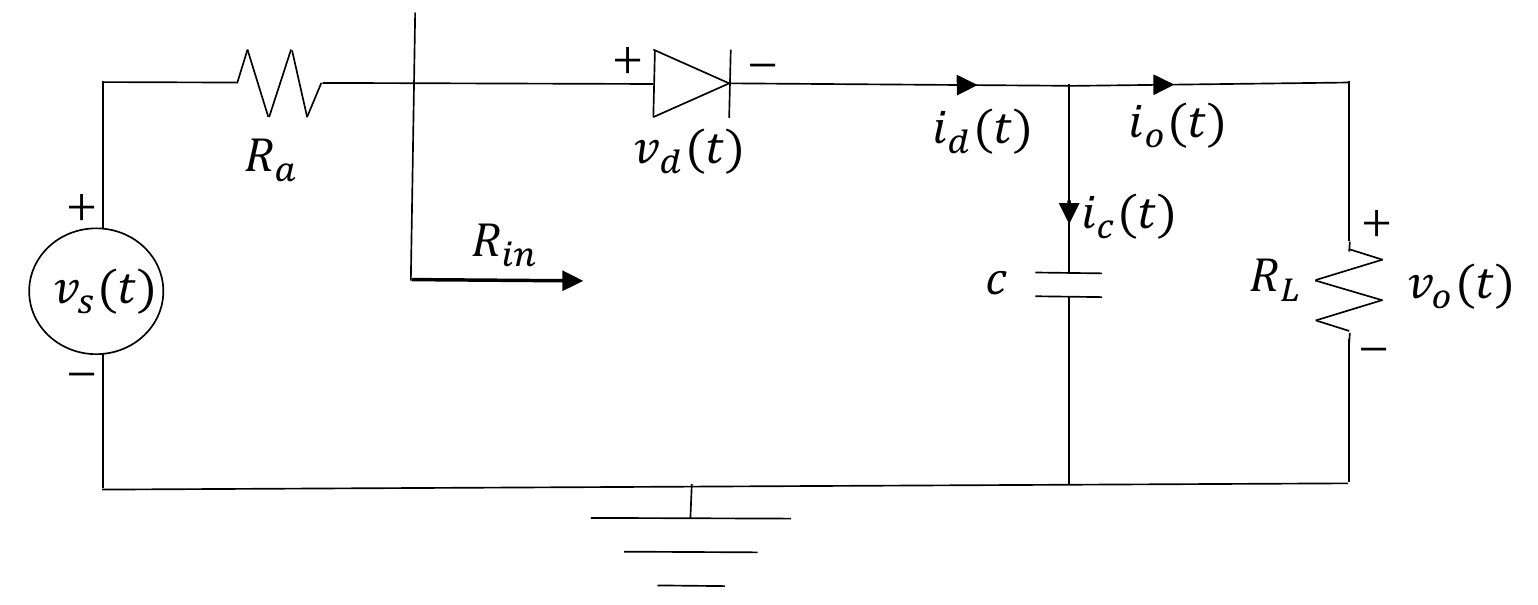}
\caption{The nonlinear model for the rectenna circuit }\label{Fig_1}
\par\end{centering}
\vspace{0mm}
\end{figure}

Adopting rectenna nonlinear model of \footnote{Rectenna is composed of an antenna and a rectifier. The antenna is modelled as a voltage source followed by a resistance and the rectifier is modelled as a nonlinear diode followed by low pass filter (LPF).} \cite{Clerckx_Bayguzina_2016,Morsi_Jamali,Vedady_Zeng}, the received RF signal $y_{\text{RF}}(t)$ is converted at the rectifier's output into a DC signal across a load resistance $R_L$. Assuming perfect impedance matching, i.e., $R_{\text{in}}=R_{a}$ ($R_{\text{in}}$ is the equivalent input impedance of the circuit observed after the antenna), the received power is completely transferred to the rectifier. Therefore, we have $\mathbb{E}[|y_{\text{RF}}(t)|^2]=\mathbb{E}[|v_{\text{in}}(t)|^2]/R_{a}$ or equivalently $v_\text{in}(t) = y_\text{RF}(t)/\sqrt{R_{a}}$ \cite{Clerckx_Bayguzina_2016}. The current $i_d(t)$ flowing through the diode is related to the voltage drop $v_d(t)$ by the Shockley diode equation $i_d(t) = i_s (\exp(\frac{v_d(t)}{\eta V_T})-1)$, where $i_s, \eta$ and $V_T$ are the diode's reverse bias saturation current, the ideality factor (typically ranging between 1 and 2) and the
thermal voltage (approximately $25.85$ mV at room temperature), respectively. Assuming that the capacitance $c$ of the LPF is sufficiently large, the output voltage can be assumed constant, i.e., $v_{o}(t) \thickapprox v_{o}$ \cite{Vedady_Zeng}. Applying Kirchoff's current law to the circuit in Figure \ref{Fig_1}, we have
\begin{align}\label{eq:6}
i_d(t)&=i_s\left(e^{\frac{v_d(t)}{\eta v_T}}-1\right)=i_s\left(e^{\frac{-v_o(t)+y_{\text{RF}}(t) \sqrt{R_{\text{a}}}}{\eta v_T}}-1\right)\\\label{eq:5}
&=i_c(t)+i_o(t)=c \frac{dv_o(t)}{dt}+\frac{v_o(t)}{R_L}=  \frac{v_o}{R_L}
\end{align}
where (\ref{eq:6}) is due to $v_d(t) = v_{\text{in}}(t)-v_o = y_{\text{RF}}(t)\sqrt{R_{a}}-v_o$, and (\ref{eq:5}) is due to $\frac{dv_o(t)}{dt}\thickapprox0$, (recall $v_{o}$ is approximately constant). Reformulating the RHS of the equations (\ref{eq:6}) and (\ref{eq:5}), and averaging (over one sumbol duration and randomness of the channel input), we obtain\footnote{We note that the nonlinear model presented in this paper is a generalization of the nonlinear model to complex channel inputs introduced in \cite{Vedady_Zeng,Morsi_Jamali}.}
\begin{align}\label{eq:1}
\mathbb{E}\left[\frac{1}{T}\int_{T}e^{By_{\text{RF}}(t)}dt\right]&=\left(1+\frac{v_o(t)}{i_sR_L}\right)e^{\frac{v_o(t)}{\eta v_T}},
\end{align}
where $B\triangleq \frac{\sqrt{R_a}}{\eta V_T}$. The DC power delivered to the load is $p_o = v_o^2/R_L$. Note that the RHS of (\ref{eq:1}) strictly increases with $v_o$. Hence, imposing a minimum delivery power constraint $p_o \geq p_{d}$ is equivalent to imposing constraint on (\ref{eq:1}), i.e.,
\begin{align}\label{eq:3}
\mathbb{E}\left[\frac{1}{T}\int_{T}e^{By_{\text{RF}}(t)}dt\right]\geq \left(1+\frac{\sqrt{p_d}}{i_s\sqrt{R_L}}\right)e^{\frac{\sqrt{R_L p_d}}{\eta v_T}}\triangleq P_{\text{del}}.
\end{align}
Assuming a rectangular pulse $g(t)$ with unit amplitude and duration $T$, we have $x(t) = x[k]$ in time slot $k$. Hence, the received signal in the RF domain reduces to $y_{\text{RF}}=\sqrt{2}B(\text{Re}\{x[k]\}\cos{2\pi f_c t}-\text{Im}\{x[k]\}\sin{2\pi f_c t})$ in time slot $k$, where the symbol $x[k]$ is a realization of random variable $X$ at time slot $k$. Hence, (\ref{eq:1}) reads as
\begin{align}\nonumber
&\mathbb{E}\left[\frac{1}{T}\int_{T}e^{By_{\text{RF}}(t)}dt\right]\\ &=\mathbb{E}\left[\frac{1}{T}\int_{T}e^{\sqrt{2}B(\text{Re}\{y(t)\}\cos{2\pi f_c t}-\text{Im}\{y(t)\}\sin{2\pi f_c t})}dt\right]\\\label{eq:7}
&\thickapprox \mathbb{E}\left[\frac{1}{T}\int_{T}e^{\sqrt{2}B(\text{Re}\{X\}\cos{2\pi f_c t}-\text{Im}\{X\}\sin{2\pi f_c t})}dt\right]\\\label{eq:2}
&= \mathbb{E}\left[I_0(\sqrt{2}B|X|)\right],
\end{align}
where in (\ref{eq:7}) we have neglected the effect of noise and in (\ref{eq:2}), $I_0(\cdot)$ is the modified Bessel function of the first kind and order zero, and the equality is due to \cite[Sec. 3.338, Eq. 4]{gradshteyn2007}. Using (\ref{eq:2}), the EH constraint reduces to
\begin{align}\label{eq:4}
\mathbb{E}\left[I_0(\sqrt{2}B|X|)\right]\geq P_{\text{del}}.
\end{align}

\section{Implementation}\label{Sec:Implemen}
We model a SWIPT system as an autoencoder, where both the transmitter and receiver are implemented as two DNNs in order to perform the encoding and decoding processes, respectively.
The transmitter communicates one of $M$ possible messages $s \in \mathcal{M} = \{1,2,...,M\}$, where each message $s$ carries $\log_2(M)$ bits, and $\mathcal{M}$ denotes the message alphabet set. In order to be transmitted, the message $s \in \mathcal{M}$ is transformed in one-hot vector (a $M$-dimensional vector of all zeros except one in $s^{\text{th}}$ position). The one-hot vector corresponding to the message $s$ is denoted by $\mathbf{s}$. The DNN maps then the vector $\mathbf{s}$ into a codeword $\mathbf{x}^{n}\in \mathcal{X}^n$ of $n$ complex symbols. The mapping from the set of messages $\mathcal{M}$ to the transmitted signal space $\mathcal{X}^n$ is denoted by $g_{\theta_T}(\cdot): \mathcal{M} \rightarrow \mathbb{C}^n$, where $\theta_T$ refers to the set of transmitter parameters, related to the weights and biases across the layers of the DNN. As the weights and the biases of the network are real numbers, each symbol of the codeword is represented by two output units corresponding to the real and imaginary part of the symbol.
We note that to satisfy the average power constraint at the transmitter, a power normalization layer is included as the last layer of the transmitter. The encoded signal $\mathbf{x}^n$ is corrupted by the channel noise (here we consider AWGN). The received signal at the receiver is denoted by $\mathbf{y}^n$.

\begin{figure}
\begin{centering}
\includegraphics[scale=0.52]{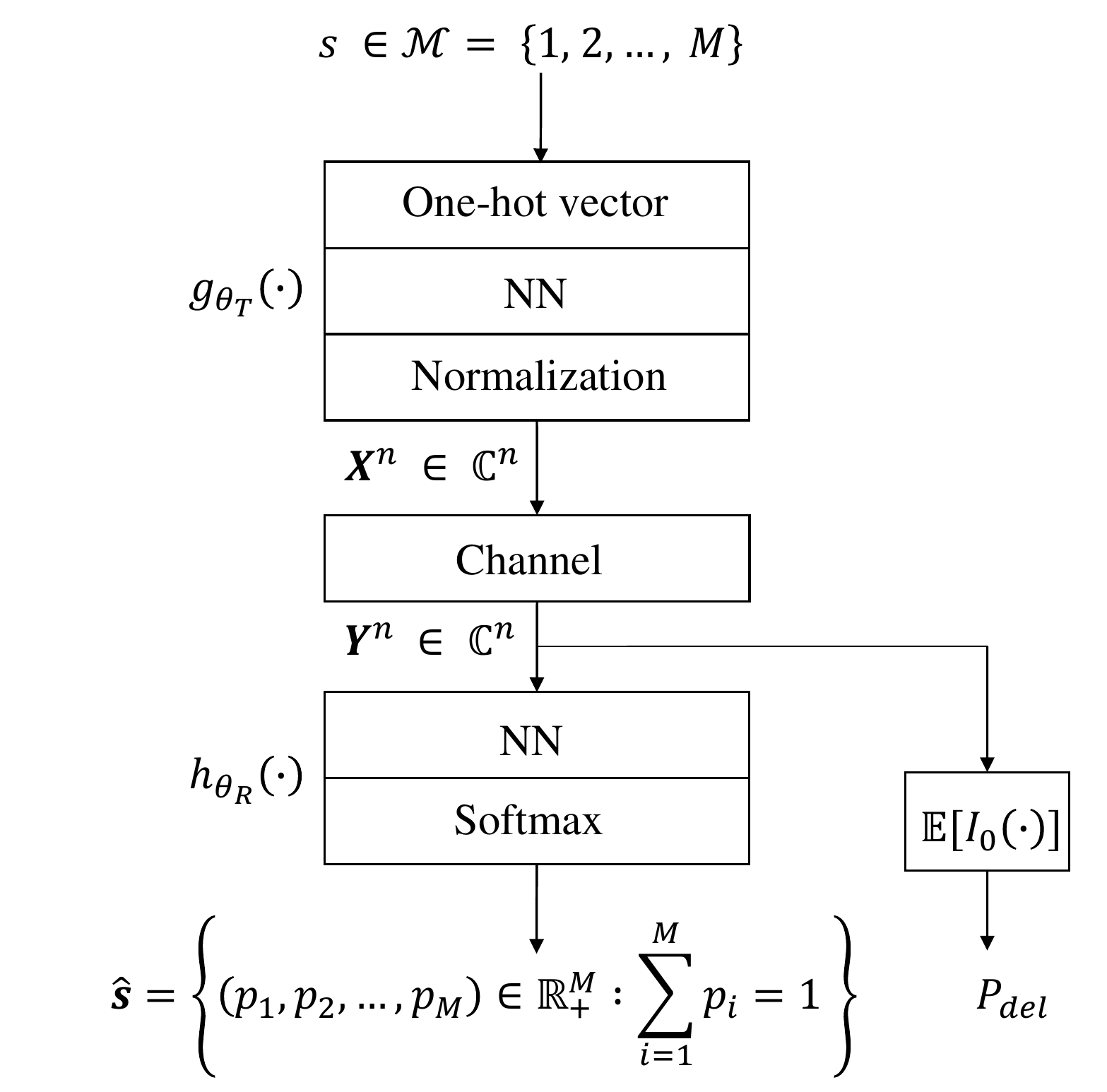}
\caption{Illustration of the autoencoder structure for the problem studied in this paper.}\label{Fig_3}
\par\end{centering}
\vspace{0mm}
\end{figure}

The receiver aims both to detect the transmitted symbol $s$ as well as harvest the delivered power $P_{\text{del}}$. The decoding is performed by mapping the received noisy codeword $\mathbf{y}^n$ to an $M$-dimensional probability vector denoted by $\hat{\mathbf{s}}$ (and outputting the detected message by obtaining the index corresponding to the maximum probability) through a parametric function defined by a fully-connected DNN $h_{\theta_R}(\cdot): \mathbb{C}^n \rightarrow \mathcal{M}$. $\theta_R$ refers to the set of receiver parameters in terms the weights and biases across different layers of the DNN to be optimized. Note that the communication rate for this system is $\log_2(M)/n$ bits per channel use.

The delivery power $P_{\text{del}}$ harvested at the receiver is modelled as in (\ref{eq:4}). We recall that for the power delivery purposes, the received RF signal is directly fed to the EH. Therefore, for power delivery purposes, the signal is not processed through the DNN.

We model the information loss as the cross entropy function between the transmitted one-hot vector $\mathbf{s}$, and the output probability vector $\mathbf{\hat{s}}$ at the receiver, i.e.,  $\mathcal{L}(\mathbf{s}, \mathbf{\hat{s}}) = \sum_{i=1}^M {\mathbf{s}_i \log \hat{\mathbf{s}}_i}$, where the $\mathbf{s}_i$ and $\hat{\mathbf{s}}_i$ indicate the $i^{\text{th}}$ entry of the vectors $\mathbf{s}$ and $\hat{\mathbf{s}}$, respectively. Accordingly, the cost function used in order to optimize the system is given by
\begin{align} \label{cost_function}
L(\theta_T,\theta_R) = \frac{1}{m}\sum_{k=0}^m \mathcal{L}(\mathbf{s}^{(k)}, \mathbf{\hat{s}}^{(k)}) +\frac{\lambda}{P_{\text{del}}},
\end{align}
where $m$ is the size of the training data, which is assumed independent and identically distributed (iid). Note that different values of the parameter $\lambda\geq 0$ in (\ref{cost_function}) can be associated to different information rate and power demands at the receiver. In our implementation of the DNN-based autoencoder, which includes the encoder at the transmitter, an AWGN channel between the transmitter and receiver and the decoder at the receiver, we have used the Adam mini-batch Gradient Descent algorithm with the programming written in Tensorflow.

\begin{figure}
\begin{centering}
\includegraphics[scale=0.33]{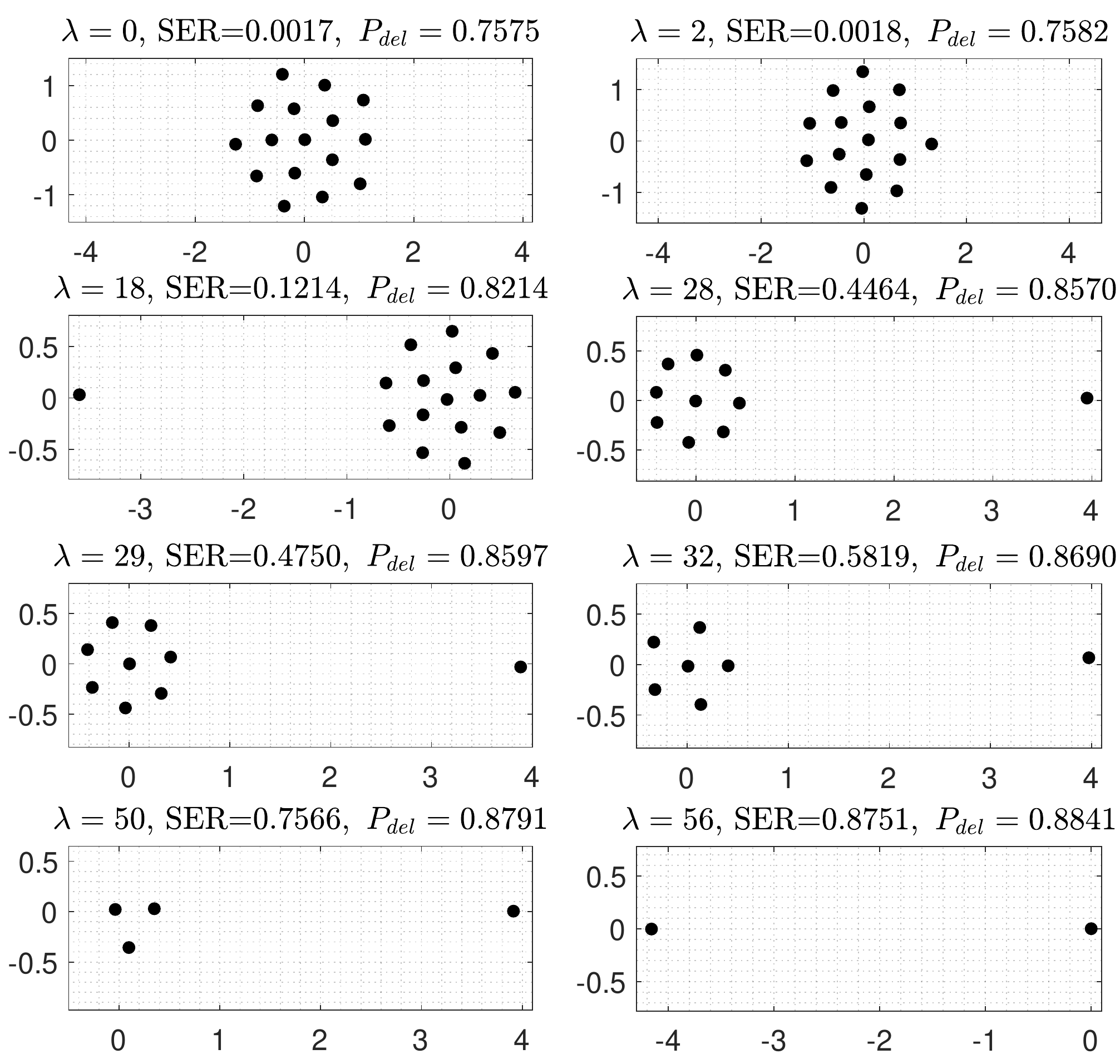}
\caption{Representation of 16-symbols modulation for different values of $\lambda$ (different information rate and power demand at the receiver) with SNR $20$ dB. By increasing $\lambda$, the delivered power at the receiver increases.}\label{Fig_4}
\par\end{centering}
\vspace{0mm}
\end{figure}

\section{Numerical Results}\label{Sec:Numerical}
In our model, we consider a training set of $m=10^5$ symbols. The training of the NN is implemented using Adam mini-batch gradient descent (MGD) algorithm with mini batch sizes of $10^3$. In order to decrease the dependency of the final solution on the initialization of the algorithm, we run the algorithm $N$ times with the same design parameters (here $N=500$) and each time with a different seed for initialization. Each message\footnote{We assume that the message set follows a uniform distribution.} is transmitted over the complex baseband channel using $n$ channel uses (in this paper we have assumes $n=1$ corresponding the communication rate $\log_2(M)$ bits per channel use). We consider a certain threshold as the maximum allowable Symbol Error Rate $\text{SER}_{\text{max}}$ (here we consider $\text{SER}_{\text{max}}=0.95$). The objective is to minimize the cost in (\ref{cost_function}) (for a predetermined size of the messages) for different values of $\lambda$, while keeping the SER of the transmission less than or equal to $\text{SER}_{\text{max}}$, i.e., SER$\leq\text{SER}_{\text{max}}$. Accordingly, for each message size, the value of $\lambda$ in (\ref{cost_function}) is increased incrementally, starting from $\lambda=0$ (Note that $\lambda=0$ is equivalent to information-only demands). We continue increasing $\lambda$ until the inequality SER$\leq\text{SER}_{\text{max}}$ is contradicted or $\lambda$ is larger than a threshold (here we consider $\lambda\leq 100$).
In Figure \ref{Fig_4}, the transmitted signal modulations are shown for $M=16$ and for different values of $\lambda$. Recall from (\ref{cost_function}) that $\lambda$ is interpreted as a factor to control the information rate and power demand at the receiver. By increasing $\lambda$, the demand for power at the receiver increases. Accordingly, the transmitted signal modulation loses its symmetry around the origin in a way that one of the transmitted symbols (power symbol) is getting away from the origin, either along the real or imaginary subchannel. This observation is similar to the result in \cite{Varasteh_Rassouli_Clerckx_ITW_2017}, where it is shown that for the Gaussian inputs, in order to have the maximum delivered power at the receiver, the transmitter is to allocate its power budget to solely real or imaginary subchannels. Another observation is that, as the power demand at the receiver increases, the transmitter sacrifices some of the messages by mapping them to the zero symbol (e.g., see the last five signal constellations in Figure \ref{Fig_4}). In the extreme scenario, where the receiver merely demands for power (still some information is transmitted over the channel, however with a very high SER), we have only two symbols (indeed one power symbol far from the origin and the remaining information symbols collapsing on top of each other at zero). Additionally, an interesting observation about the SWIPT modulations in Figure \ref{Fig_4} (specifically focusing on the last modulation and considering a very long transmission) is that, they approach to distributions with low-probability/high-amplitudes and high-probability/zero-amplitudes. This result is also in line with the result obtained in \cite{Varasteh_Rassouli_Clerckx_arxiv}, where it is shown that the optimal channel input distributions for power delivery purposes (accounting for nonlinearity with some simplification assumptions) follow the same behaviour, i.e., low probability-high amplitudes and high probability-zero amplitudes.

\begin{figure}
\begin{centering}
\includegraphics[scale=0.3]{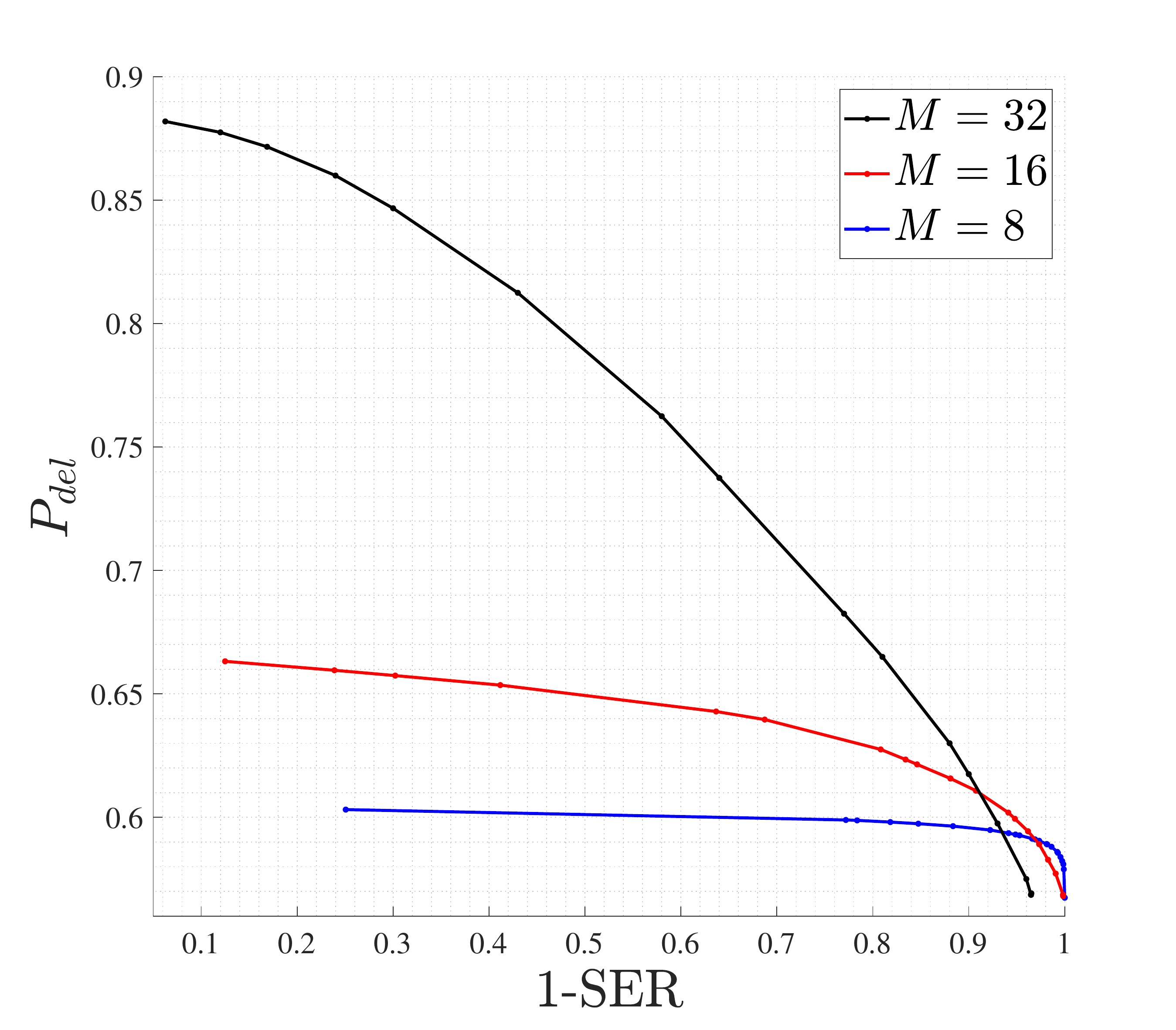}
\caption{Representation of the tradeoff between the delivered power and information rate at the receiver. The delivered power at the receiver increases with the number of symbols of the transmitted signal constellation.}\label{Fig_5}
\par\end{centering}
\vspace{0mm}
\end{figure}

In Figure \ref{Fig_5}, the delivered power $P_{\text{del}}$ versus complementary of symbol error rate ($1-\text{SER}$) for different message sizes ($M=8,~16,~32$) with signal to noise ratio $\text{SNR}=20$dB is illustrated. It is observed that as the size of the message is increased (which is equivalent to increasing the channel input symbols), the delivered power at the receiver increases as well. This can be justified as follows. For power delivery purposes, the transmitter favours distributions with high probability information symbols around zero and a low probability power symbol away from zero. Noting that the message set is uniformly distributed, such a distribution can be achieved by having more symbols around zero and one symbol away from zero. It can be easily verified that the probability of the power symbol (equivalently the occurrence of the power symbol in the long term) and its amplitude decreases and increases, respectively, with the size of the message set. This in turn results in more delivered power at the receiver. As the last point, we note that, due to the nonlineary effect (dependency of the delivered power on Bessel function in (\ref{eq:3})), the delivered power is directly dependent on the channel input average input power constraint. This is equivalent to the fact that two systems with the same SNR but different average power levels result in different designs. Due to lack of space, we have postponed this investigation for the longer version of the paper.

\section{Conclusion}\label{Sec:Concl}
In this paper, we studied a point-to-point SWIPT signal and system design. We considered the system as an autoencoder, where the transmitter and the receiver are implemented as deep neural networks. The end-to-end optimization of the system is done by jointly learning the transmitter and receiver parameters as well as signal encoding. We considered the case where the transmitter uses one complex symbol to transmit each message. The numerical results reveal that, as the power demand at the receiver increases, the transmitted signal modulation is reshaped, such that one of the symbols (power symbol) is shot away from the origin along either real or imaginary subchannel and the other symbols (information symbols) are symmetrically distributed around the origin. As future research directions, we note that short block length transmissions as well as obtaining a model that features the practical limitations of the rectenna (nonlinearity) accurately, are under investigation.

\bibliographystyle{IEEEbib}
\bibliography{strings,ref}

\end{document}